\documentstyle[aps]{revtex}
\input{epsf.tex}                 

\begin{document}
\draft
\title{Collective modes of a trapped Lieb-Liniger gas:\\ 
a hydrodynamic approach}
\author{J. N. Fuchs}
\address{Laboratoire Kastler Brossel,
Ecole Normale Sup\'erieure,\\
24 rue Lhomond, 75231 Paris Cedex 05, France \\
email: fuchs@lkb.ens.fr}
\author{X. Leyronas and R. Combescot}
\address{Laboratoire de Physique Statistique,
Ecole Normale Sup\'erieure,\\
24 rue Lhomond, 75231 Paris Cedex 05, France}
\date{Received \today}
\maketitle

\begin{abstract}
We consider a trapped repulsive one-dimensional (1D) Bose 
gas at very low temperature. In order to study the collective modes of this 
strongly interacting system, we use a hydrodynamic approach, where the gas 
is locally described by the Lieb-Liniger model of bosons interacting 
via a repulsive delta potential. Solving the corresponding linearized 
hydrodynamic equations, we obtain the collective modes and concentrate more
specifically on the lowest compressional mode. This is done by finding
models, approaching very closely the exact equation of state of the gas, for
which the linearized hydrodynamic equations are exactly solvable. Results
are in excellent agreement with those of the sum rule approach of Menotti
and Stringari.
%
%
%
\end{abstract}


\section{INTRODUCTION}

Cold atomic gases of reduced dimensionality are currently being investigated
experimentally \cite{Schreck,Gorlitz,Zurich}. A typical situation is that of a dilute 
3D Bose-Einstein condensate (BEC) confined in a very anisotropic trap. If the 
confinement is strong enough in two perpendicular directions, the gas behaves 
microscopically as a 1D Bose gas in the remaining perpendicular direction. A very 
interesting model of such an interacting system is that of Lieb and Liniger 
\cite{Lieb}: it consists of 1D bosons interacting via repulsive zero-range interactions 
of variable strength. In the \emph{homogeneous} case, this model can be solved exactly, 
whatever the strength of the interactions, by means of Bethe ansatz \cite{Lieb}.

In the present paper, we propose to study the lowest lying collective modes of the 
\emph{trapped} Lieb-Liniger gas by means of a hydrodynamic approach \cite{CL,MS}. The 
linearized hydrodynamic equations are solved using the recently developed method of 
Combescot and Leyronas \cite{CL}. The results, already presented in \cite{FLC}, are 
compared with those of Menotti and Stringari \cite{MS}, which were obtained from 
a sum rule approach.

\section{TRAPPED LIEB-LINIGER GAS}

The physical system we consider is a 3D dilute BEC at very low temperature ($T\simeq 0$) confined in an
elongated cigar-shaped trap. The trap is taken to be harmonic with radial frequency 
$\omega_{\perp}=\omega_{x}=\omega_{y}$ much greater than the axial
frequency $\omega_{z}$. At very low temperature, repulsive interactions
between atoms are fully characterized by the 3D s-wave scattering length $%
a>0 $. 

The temperature $T$ and the chemical potential $\mu$ are much smaller
than the radial quantum of oscillation $\hbar \omega_{\perp}$, so that the radial
motion is frozen and the gas behaves microscopically as a 1D system along
the axial direction of the trap. Olshanii \cite{Olshanii} showed that in
that case the corresponding effective 1D scattering length $a_{1}$ is given
by $a_{1}=a_{\perp}^2/a$ \cite{sign} when $a_{\perp}\equiv\sqrt{%
\hbar/m\omega_{\perp}} \gg a$, which we assume for simplicity.
In order to study such a trapped 1D Bose gas, we assume that it 
is locally described by the Lieb-Liniger model 
\cite{Lieb} of hard-core bosons, of mass $m$, interacting trough a repulsive
contact potential $g_{1}\delta(z)$, where $g_{1}=2\hbar^2/ma_{1}$ \cite
{Olshanii}.

In the Lieb-Liniger model, the relevant parameter measuring
the strength of the interactions is the dimensionless coupling constant 
$\gamma\equiv mg_{1}/\hbar^2 n_{1}=2/n_{1}a_{1}$, where $n_{1}$ is the 1D density. It is
the only dimensionless intensive parameter proportional to the coupling constant 
$g_{1} $ \cite{Lieb}. It shows that the gas is 
weakly interacting $\gamma \ll 1$ in the dense limit $n_{1}a_{1} \gg 1$ 
and strongly interacting $\gamma \gg 1$ in the dilute limit $n_{1}a_{1}\ll 1$. 

In the dense limit, the homogeneous gas is a quasi-condensate, i.e. a condensate
with a fluctuating phase. Nevertheless, in a trapped gas, when the 
temperature is well below the degeneracy temperature \cite{Petrov}, 
the phase is almost constant over the whole 
sample and the gas behaves as a true condensate: it is, for example, well described 
by a 1D Gross-Pitaevskii equation \cite{GP}. In this limit, we
say that the gas is in the ``1D mean-field'' regime and the chemical
potential is given by $\mu=g_{1}n_{1}=2\hbar \omega_{\perp} a n_{1}$ (with
the convention that $\mu=0$ when $n_{1}=0$). The low density limit
corresponds to the ``Tonks-Girardeau'' regime, i.e. to a gas of impenetrable
bosons. Girardeau \cite{Girardeau} showed that in this limit there is an
exact mapping between the gas of impenetrable bosons and the 1D free 
Fermi gas. In the Tonks-Girardeau regime, the chemical potential is given by 
$\mu=(\hbar \pi n_{1})^2/2m$. In the whole range between the 1D mean-field
and the Tonks-Girardeau regime, the equation of state $\mu(n_{1})$ can be
obtained from the exact solution of the homogeneous Lieb-Liniger model \cite{Lieb}.
Menotti and Stringari numerically evaluated this equation of state and made
their result available \cite{MS,Chiara}. As we will see, the equation of
state is the essential ingredient needed in the hydrodynamic approach in
order to compute the collective modes.

In the following, we use reactive hydrodynamics, where dissipation is negligible 
and thermal effects can be omitted. This is valid at low enough 
temperature. The hydrodynamic equations are: 
\begin{eqnarray}
\frac{\partial n_{1}}{\partial t} &+& \partial_{z} (n_{1}v)=0
\label{continuity} \\
m\frac{d v}{d t} &=& -\partial_{z} \big[ \mu(n_{1})+m \omega_{z}^2 z^2/2 %
\big]  \label{Euler}
\end{eqnarray}
where $n_{1}(z,t)$ is the 1D density at position $z$ and time $t$ and $%
v(z,t) $ is the velocity along the $z$ axis. The first equation is the
continuity equation, the second is the Euler equation at zero temperature.
The equation of state for the homogeneous gas $\mu(n_{1})$ appears as a
needed input in the hydrodynamic equations: it is provided by the Lieb-Liniger
microscopic theory.

Here we briefly discuss the validity of the hydrodynamic approach. First, in
the 1D mean-field regime, a trapped gas with a finite number of particles
behaves as a true condensate at low enough temperature \cite{Petrov}. 
It is therefore well
described by a 1D Gross-Pitaevskii equation. As in the 3D case \cite{dgps},
neglecting the quantum pressure (which is valid when the number of particle
is large), the Gross-Pitaevskii equation can be rewritten in the form of
reactive hydrodynamics with the equation of state given by the mean-field
result $\mu(n_{1})=g_{1}n_{1}$. Second, for the whole range of densities 
(i.e. of coupling constant $\gamma$), the low lying excitations of the 
homogeneous Lieb-Liniger gas are known to be phonon-like with a sound 
velocity depending on the coupling constant \cite{Lieb}. 
This property allows the description of the system in term of
an effective Luttinger liquid theory \cite{Haldane}. In the case of the
trapped gas, when studying collective modes of long wavelength and low
energy, the chemical potential is locally given by its value in the
homogeneous gas at the same density $\mu[n_{1}(z)]$. This allows to
extend the hydrodynamic approach initially developed for condensates in 
the mean-field regime \cite{dgps} to the strongly interacting 
regime \cite{CL,MS}.

\section{COLLECTIVE MODES : SOLVABLE MODELS FOR HYDRODYNAMICS}

In this section we use the method recently developed by Combescot and Leyronas 
\cite{CL} in order to solve the linearized hydrodynamic equations. The basic
idea is to replace the equation of state $\mu(n_{1})$ by a model $%
\mu_{mod}(n_{1})$ for which the linearized hydrodynamic equations can be
solved analytically or quasi-analytically. The model has several parameters
that allows to fit the actual equation of state as closely as possible.
Details of the fitting procedure are given in \cite{FLC}. We start by
briefly reviewing the method of Combescot and Leyronas in the 1D case.

At equilibrium the particle density $n_{1}^0(z) $ satisfies:
\begin{equation}
\mu [n_{1}^0(z)] + m\omega_{z}^2z^2/2 = \tilde{\mu}  \label{equilibrium}
\end{equation}
where $\tilde{\mu }$ is the constant value over the system of the global
chemical potential. For small density fluctuations $\delta
n_{1}(z,t)=n_{1}(z,t)-n_{1}^0(z)$ we introduce the departure of the chemical
potential from its equilibrium value $w(z,t) = \tilde{\mu }(z,t)-\tilde{\mu}%
= (\partial \mu/\partial n_{1}^0) \delta n_{1}(z,t)$. For a fluctuation
occurring at frequency $\omega $, the preceding equation gives $i\omega
\delta n_{1}=\partial_{z}(n_{1}^0 v) =n_{1}^0\partial_{z}v
+v\partial_{z}(n_{1}^0)$ which, together with Euler equation (\ref{Euler})
leads to: 
\begin{equation}
n_{1}^0\partial_{z}^{2} w+ \partial_{z}n_{1}^0\partial_{z}w
+m\omega^{2}(\partial n_{1}^0 / \partial \mu) w=0  \label{eq1}
\end{equation}
The equilibrium relation (\ref{equilibrium}) implies $(\partial \mu/\partial
n_{1}^0)(\partial n_{1}^0 / \partial z)=-m\omega_{z}^2z$ which gives: 
\begin{equation}
zw^{\prime\prime}+ zL^{\prime}(z)w^{\prime}-\nu^2 L^{\prime}(z) w = 0
\label{eq2}
\end{equation}
where $\nu^2\equiv \omega^2/\omega_{z}^2$ is the squared
reduced mode frequency and we defined $L(z) \equiv \ln [n_{1}^0(z)]$ with 
$L^{\prime}(z) \equiv dL/dz $. Let us introduce the function $v(z)$ by 
$w(z)=z^{l}v(z)$, with $l=0$ for an even mode with respect to $z$ 
and $l=1$ for an odd mode, and restrict to $z>0$. With
this change of function, equation (\ref{eq2}) becomes: 
\begin{equation}
zv^{\prime\prime}+ [ 2l+ zL^{\prime}(z)] v^{\prime}-(\nu^2-l) L^{\prime}(z)
v = 0  \label{eq4}
\end{equation}
This equation is invariant under the replacement $z \rightarrow z/R $. We
therefore take the cloud radius $R$ as unity in the following. The cloud
radius is given by $\mu[n_{1}^0(0)]=m\omega_{z}^2 R^2/2$. Next we make the
change of variable $y=z^{\alpha}$ to obtain: 
\begin{equation}
y \frac{d ^{2}v}{ dy ^{2}} + (\Delta + y \frac{dL}{ dy}) \frac{dv}{ dy} - 
\frac{\nu ^{2} - l }{\alpha } \frac{dL}{ dy} v = 0  \label{eq6}
\end{equation}
where $\Delta = 1+ (2l-1)/\alpha$. Finally, we define the reduced density $%
\bar{n}_{1}(z)\equiv n_{1}(z)/n_{1}(0)$ and the normalized local chemical
potential $\bar{\mu }(z) \equiv \mu [n_{1}(z)]/\mu [n_{1}(0)]=1-z^{2}$. We
note that in the mode equation (\ref{eq6}), the equation of state only enters
trough the quantity $dL/dy$.

The simplest model introduced by Combescot and Leyronas is the two parameter 
$\alpha-p$ model \cite{CL}. It consists in replacing the quantity $dL/dy$ by
the model $-p/(1-y)$ in equation (\ref{eq6}), which then reduces to
the hypergeometric differential equation: 
\begin{equation}
y (1-y)\frac{d^{2}v}{dy^{2}} + [\Delta-y(p+\Delta)] \frac{dv}{ dy}+ p\frac{%
\nu^{2}-l}{\alpha}v=0  \label{eqhyper}
\end{equation}
The solution to this equation (with proper boundary conditions) is a 
polynomial \cite{CL}. The corresponding mode frequencies are: 
\begin{equation}
\nu^2 =l+\frac{\alpha}{p} \: n \: (n+p+\frac{2l-1}{\alpha})  \label{eqfreq}
\end{equation}
where $n$ is a positive integer and $l=0$ or $1$. The $\alpha-p$ model
corresponds explicitely to the equilibrium density 
$\bar{n}_{1}(z)=(1-z^\alpha)^p$ and to the equation of state 
$\bar{\mu}=1-(1-\bar{n}_{1}^{1/p})^{2/\alpha}$.
We can check on equation (\ref{eqfreq}) that the
``dipole mode'' ($n=0$ and $l=1$) occurs at frequency $\omega=\omega_{z}$ in
agreement with Kohn's theorem. The particular case $\alpha=2$ corresponds to
the power law equation of state $\bar{\mu }=\bar{n}_{1}^{1/p}$. This is precisely the
functional dependence of the equation of state in the two limiting regimes
that we discussed above: the Tonks-Girardeau regime corresponds to $p=1/2$ and
the 1D mean-field regime to $p=1$ . In the Tonks-Girardeau case, the
mode frequencies are given by $\omega=k\omega_{z}$, where $k\equiv2n+l$ is
any positive integer (it is convenient to notice that $l^2=l$). 
This result was already obtained in \cite{Minguzzi,MS}. 
In the 1D mean field regime, the mode frequencies are given by 
$\nu^2=k(1+k)/2$, where $k\equiv2n+l$ is any positive integer, in agreement
with \cite{MS}.

Combescot and Leyronas introduced other models which lead to a generalized
hypergeometric differential equation with quasi-polynomial solutions \cite
{CL}. Here, we only describe the ``3 parameters quasi-polynomial'' model. The
quantity $dL/dy$ is modeled by $-(p_{0}+p_{1}y)/(1-y)$ depending on the
three parameters $\alpha$, $p_{0}$ and $p_{1}$. It is solution of: 
\begin{equation}
y(1-y)\frac{d ^{2}v}{dy^{2}}+(-p_{1}y^{2}-(\Delta+p _{0})y + \Delta) \frac{dv%
}{ dy} + \frac{\nu^2-l}{\alpha}(p_{1}y +p_{0}) v = 0  \label{equasipol}
\end{equation}
This differential equation admits solutions that are very rapidly converging 
series which can safely be truncated above some order and that we call 
quasi-polynomials \cite{CL}. The mode frequencies are easily obtained 
numerically. Details are given in \cite{FLC}.

From now on, we will concentrate on the lowest compressional mode
($n=1$ and $l=0$). In order to obtain the mode frequency,
we will use the following procedure: (i) fit the actual equation of state by
a model (either $\alpha-p$ or 3 parameters quasi-polynomial) to obtain the
best set of parameters; (ii) calculate the mode frequency using this set of
parameters (with equation (\ref{eqfreq}) in the case of the $\alpha-p$ model,
for example). In the next section, we show how to improve on this zero-order
result for the mode frequency by using perturbation theory.

\section{PERTURBATION THEORY}

In order to account for the difference between the actual equation of state
and the model equation of state we can rewrite the mode equation (\ref{eq6})
in the form of a Schr\"{o}dinger-like equation and use standard perturbation
theory. This allows to correct the results obtained for the mode frequency
calculated using the model equation of state and to come closer to the mode
frequency corresponding to the actual equation of state.

The following change of function $\psi(y)=v(y)\big[y^\Delta n_{1}^0(y)\big]$
is made in equation (\ref{eq6}) leading to: 
\begin{eqnarray}
-\frac{d ^{2} \psi }{ dy ^{2}} + \Big[\frac{1}{2} L^{\prime\prime}+ \frac{1}{%
4} L^{^{\prime}2} + (\frac{\nu ^{2} - l }{\alpha }+ \frac{ \Delta }{2} ) 
\frac{ L^{\prime}}{y}+ \frac{ \Delta ( \Delta - 2)}{4 y ^{2}}\Big] \psi = 0
\label{eq7t}
\end{eqnarray}
which is a Schr\"{o}dinger equation with an effective potential 
(term between brackets $[$ $]$) and
corresponding to zero energy (and $\hbar ^{2}/2m = 1$). The effective
potential can be decomposed into a zero-order term corresponding to the
model equation of state and a perturbation term corresponding to the
difference of effective potential between the actual and the model equation
of state. The perturbation is treated using standard first order
perturbation theory. This leads to a first order change of the energy in the
Schr\"{o}dinger equation. In order to keep the energy equal to zero, we have
to give a compensating variation $\delta \nu ^{2}$ for the frequency: 
\begin{equation}
\delta \nu ^{2}=\alpha \frac{\int_{0}^{1} \! dy \: \delta L^{\prime}\, n
_{1}^0 \, y ^{\Delta -1} v [\frac{\nu ^{2} - l }{\alpha } v - y v^{\prime}]%
} {\int_{0}^{1} \! dy \: (-n_{1}^{0\prime}) \, y ^{\Delta -1}v^{2}}
\label{eqcorr}
\end{equation}
where $\delta L^{\prime}\equiv L^{\prime}-L_{mod}^{\prime}$. This result is
proven in \cite{FLC}.

\section{LOWEST COMPRESSIONAL MODE}

We apply now the previously developed method to study the lowest
compressional mode ($n=1$ and $l=0$), which physically corresponds 
to a breathing motion of the gas along the axis. The needed input is
the equation of state $\mu(n_{1})$ of the Lieb-Liniger model. Using 
the numerical evaluation of this equation of state 
by Menotti and Stringari, we compute the mode frequency as follows:
for each value of the dimensionless density at the center of the trap 
$n_1(0)a_{1}$, (i) we fit the equation of state (for $n_1(z)$ varying between $0$ and 
$n_1(0)$) with either the $\alpha-p$ or the 3 parameters quasi-polynomial
models; (ii) the zero order mode frequency is then obtained by inserting the
value of the best set of parameters in the formula giving the mode frequency
for the model; (iii) we then compute the first order perturbation correction to
the mode frequency with equation (\ref{eqcorr}).

\begin{figure}
\centering
\vbox to 70mm{\hspace{1mm} \epsfysize=7cm 
\epsfbox{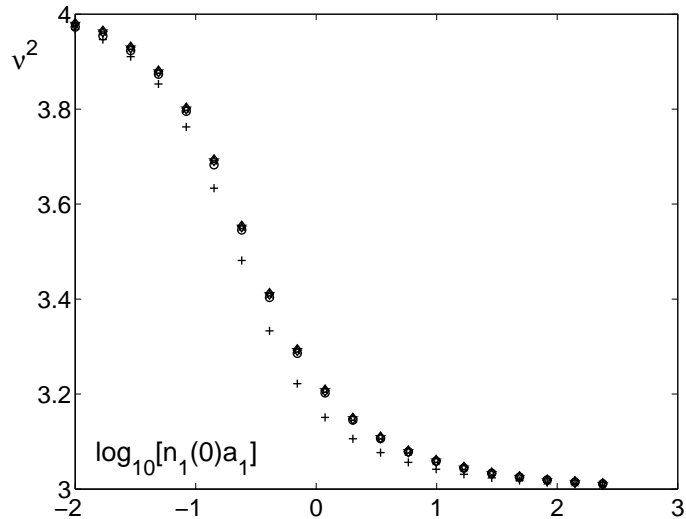} }
\caption{Lowest compressional mode of the trapped Lieb-Liniger gas. The squared
mode frequency $\protect\nu^2=\protect\omega^2/\protect\omega_{z}^2$ is
plotted as a function of $\log_{10}[n_{1}(0)a_{1}]$. The crosses ($+$)
correspond to the $\protect\alpha-p$ model; the stars ($\star$) to the
corrected $\protect\alpha-p$ model; the circles ($\circ$) to the 3
parameters quasi-polynomial model; and the diamonds ($\diamond$) to the
corrected 3 parameters quasi-polynomial model.}
\label{nu2pro}
\end{figure}

Figure \ref{nu2pro} shows the results of four different calculations of the
lowest compressional mode using respectively the $\alpha-p$ model and the 3 parameters
quasi-polynomial model without perturbative correction, 
and the same models corrected by first order
perturbation theory. Actually, at this scale, it is hard to see any
difference between these calculations, except for the $\alpha-p$ model which
gives a slightly lower mode frequency. 

\begin{figure}
\centering
\vbox to 70mm{\hspace{1mm} \epsfysize=7cm 
\epsfbox{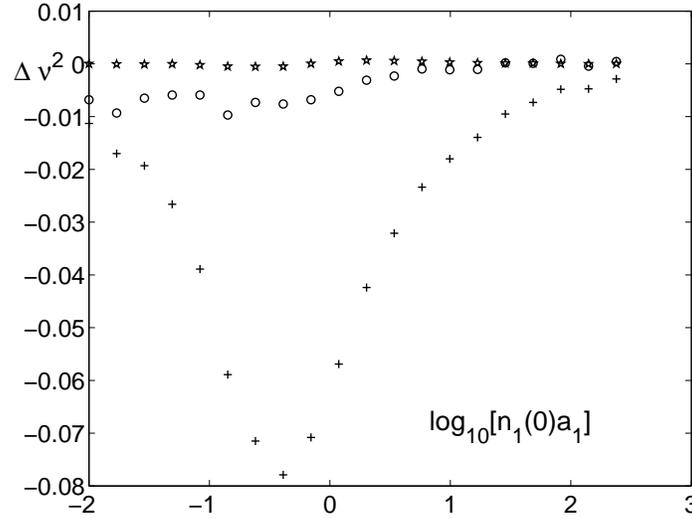} }
\caption{Lowest compressional mode of the trapped Lieb-Liniger gas. The squared
mode frequency $\protect\nu^2=\protect\omega^2/\protect\omega_{z}^2$
calculated using the corrected 3 parameters quasi-polynomial model is taken
as a reference. The difference between this reference and three different
calculations ($\protect\alpha-p$, corrected $\protect\alpha-p$ and 3
parameters quasi-polynomial models) are plotted as a function of $%
\log_{10}[n_{1}(0)a_{1}]$. The crosses ($+$) correspond to the $\protect%
\alpha-p$ model; the stars ($\star$) to the corrected $\protect\alpha-p$
model; and the circles ($\circ$) to the 3 parameters quasi-polynomial model.}
\label{diffpro}
\end{figure}

In order to compare the various calculations, we plot the same data in a
different manner, see figure \ref{diffpro}. As discussed in \cite{FLC} we
expect the corrected 3 parameters model to be the most precise of our results.
We therefore take this corrected model as a reference and plot the differences 
between the mode frequencies calculated within the three other approaches 
($\alpha-p$ , corrected $\alpha-p$ and 3 parameters
quasi-polynomial models) and this reference, as shown in figure \ref{diffpro}. 
We can see on this figure that, once corrected, the results of the 
$\alpha-p$ and the 3 parameters quasi-polynomial models agree remarkably well, at the
absolute $10^{-3}$ level for $\nu^2$. As already mentioned, the $\alpha-p$ model
is exact in the 1D mean-field and in the Tonks-Girardeau limits where it
predicts $\nu^2=3$ and $\nu^2=4$, respectively. In between these limits, the 
uncorrected $\alpha-p$ model has some difficulties in predicting precisely the correct
value of the mode frequency. For a discussion of this issue see \cite{FLC}.

\begin{figure}
\centering
\vbox to 70mm{\hspace{1mm} \epsfysize=7cm 
\epsfbox{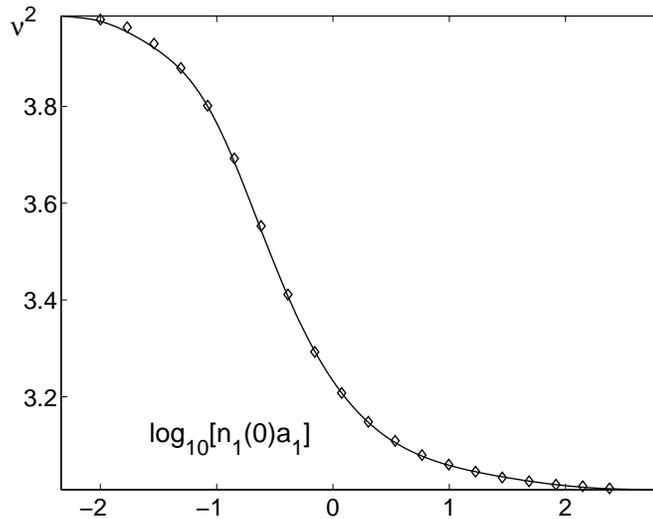} }
\caption{Lowest compressional mode of the trapped Lieb-Liniger gas. The squared
mode frequency $\protect\nu^2=\protect\omega^2/\protect\omega_{z}^2$ is
plotted as a function of $\log_{10}[n_{1}(0)a_{1}]$. The diamonds ($\diamond$%
) correspond to the corrected 3 parameters quasi-polynomial model, and the
full line is the result of Menotti and Stringari.}
\label{nu2mspro}
\end{figure}

Next we want to compare our most precise result (the corrected 3 parameters
quasi-polynomial model) with the results of Menotti and Stringari \cite{MS},
who calculated the lowest compressional mode frequency using a sum rule
technique \cite{dgps}, which is basically a variational approach. 
This method is known to provide an exact upper bound to the lowest 
mode frequency. Both results are shown in figure \ref{nu2mspro}. It is seen
that they agree very well over the whole range of $n_1 (0)a_{1}$. 

\section{DISCUSSION}
In this paper, we used a recently developed method to solve the linearized 
hydrodynamic equations of the trapped Lieb-Liniger gas at $T \simeq 0$. 
We obtained the collective modes with high precision and especially studied 
the lowest compressional mode which is relevant to current experiments. 
Moritz et al. \cite{Zurich} recently measured this mode in a trapped 1D Bose 
gas in the mean-field regime. They found $\nu^2\simeq 3$ when the gas is 
degenerate, in agreement with our calculations. As first emphasized in \cite{MS}, 
measuring the lowest compressional mode should allow to reveal the transition 
from the 1D mean-field to the Tonks-Girardeau regime, where fermionization is 
expected.

More generally, we showed, in the particular case of the trapped Lieb-Liniger gas, 
how the hydrodynamic approach allows to obtain collective modes even in a strongly 
interacting system. We also checked that the sum rule approach used by Menotti and 
Stringari \cite{MS} indeed gives very good results for the trapped 1D Bose gas. For 
a discussion of a system where this is not necessarily true see \cite{FLC}.

\section*{ACKNOWLEDGMENTS}

We are grateful to Sandro Stringari, Chiara Menotti and Dima Gangardt for many
stimulating discussions. Laboratoire Kastler Brossel and Laboratoire de 
Physique Statistique are Unit\'{e}s de Recherche de l'\'Ecole Normale 
Sup\'erieure (ENS) 
et des Universit\'es Paris 6 et Paris 7, associ\'ees au Centre 
National de la Recherche Scientifique (CNRS).


\begin{references}
\bibitem{sign}  We use a definition of the effective 1D scattering length 
similar to that of Menotti and Stringari \cite{MS}. There is a difference 
in sign with the definition of Olshanii \cite{Olshanii}.
\end{references}
\end{document}